\documentclass[reprint,amsmath,amssymb,aps]{revtex4-1}
\pdfoutput=1

\usepackage{graphicx}
\usepackage{dcolumn}
\usepackage{bm}

\begin{document}

\preprint{APS/123-QED}

\title{Gravitational waves detection with exceptional points in micro cavities}
\thanks{}%

\author{Jian Liu}
 \altaffiliation[]{}
 \author{Lei Chen}
 \altaffiliation[]{}
  \author{Fei He}
 \altaffiliation[]{}
\author{Ka-Di Zhu}%
 \email{zhukadi@sjtu.edu.cn}
\affiliation{Key Laboratory of Artificial Structures and Quantum Control (Ministry of
Education), School of Physics and Astronomy, Shanghai Jiao Tong
University, 800 DongChuan Road, Shanghai 200240, China,
Collaborative Innovation Center of Advanced Microstructures, Nanjing, China
}%

\date{\today}

\begin{abstract}
Here we propose a new gravitational waves(GWs) detector in broad frequency
band, which is operated at exceptional points(EPs) in micro cavities.
The detected signal is an eigenfrequency split of the mechanical
modes caused by the spatial strain. Due to the complex square root topology
near the EP, the splitting is greatly enhanced for sufficiently small
perturbations. Compared to current strategies, it can be achieved at the
room temperature and has advantages in micro device scale, wide frequency
band and higher sensitivity.

\end{abstract}

\pacs{Valid PACS appear here}
\maketitle


\section{INTRODUCTION}

According to the quantum mechanical perturbation theory, a perturbation with
strength of $\varepsilon $ acting on a two fold-degenerate system will cause
the energy shift or energy split in proportion to $\varepsilon $. Thus it is
possible to probe variety of parameters in a system by the frequency
spectrum and some sensors are based on this principle[1-4]. There is another
type of degeneracy in the open system (exchanging energy with the
surrounding environment), called exceptional point (EP). The EP is the
spectrum singularity in the parameter space, where two or more eigenvalues
and their corresponding eigenvectors coalesce simultaneously. One of the
main differences between exception points and conventional degeneration is
their sensitivity to perturbations[5-7]. Owing to the complex square-root
topology near an EP, any perturbation $\varepsilon $ lifts the degeneracy,
leading to a frequency splitting that scales as $\sqrt{\varepsilon }$.
Therefore, a smaller perturbation means that the improvement in sensitivity
is more significant.

As well known from the general theory of relativity, gravitational
waves(GWs) can cause a strain $h$ in space. Existing detection strategies
are based on long-baseline optical interferometry [8]. The principle is to
utilize the time-varying phase shifts caused by GWs in the optical path.
Since the wavelengths or frequencies of radiated GWs are determined by the
scales of the sources, it is necessary to observe the GWs with various
frequency bands to understand the hierarchical structure of the universe.
At present, laser interferometer detectors have been improved,
and it is expected to directly detect GWs in the 0.1 to 1 kHz band
through advanced LIGO technology[9]. Observations of low-frequency GWs have
been tried through spaceborne experiments[10-12] and astrophysical
observations[13].

In this Letter, we study how micro scale optical cavities can be used to
detect gravitational wave radiations. We propose a new GW
detection mechanism based on the optomechanical coupling change induced by
spatial strain at the EP of the system. This scheme does not rely on a
shot-noise limited displacement measurement of test mass mirrors, but rather
depends on a precision frequency measurement of the nanomechanical
resonators. The eigenfrequency splitting induced by GW can be read out in
the high-resolution frequency spectrum. The proposed detector differs from
known detectors in at least 4 points: (i) the scale of the device is not
limited, here we use the micro cavities for example. (ii) can detect
gravitational waves over a wider frequency range. (iii) due to the
topological properties at the EP, the sensitivity is greatly improved. (iv)
operate at room temperature.

\section{THEORY FRAMEWORK}

\begin{figure*}[tbp]
\includegraphics[width=12cm]{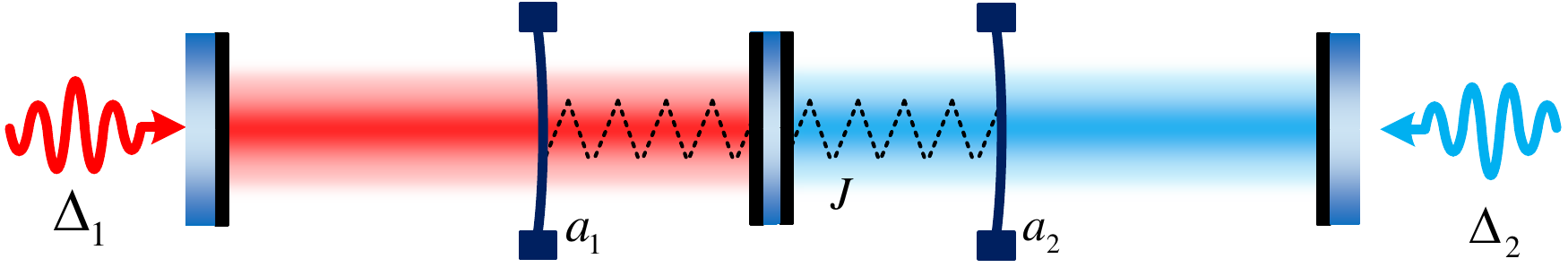}
\caption{ Generic setup. An open system consisted of coupled double
resonators . The two resonators are coupled to each other, and each
resonator is coupled to an optical cavity.The two cavities are driven by
blue- and red-detuned laser, respectively.}
\end{figure*}
Let us begin with the Hamiltonian of coupled mechanical resonators,

\begin{equation}
H=\hbar \omega _{1}a_{1}^{+}a_{1}+\hbar \omega _{2}a_{2}^{+}a_{2}+\hbar
J(a_{1}^{+}a_{2}+a_{1}a_{2}^{+}).
\end{equation}

Here $a_{1}^{+}(a_{1})$ and $a_{2}^{+}(a_{2})$ are the creation
(annihilation) operators of the mechanical modes, $\omega _{1}$ and $\omega
_{2}$ are their mechanical frequencies, $J$ denotes the coupling strength
between them. The tunable coupling of two mechanical resonators can be
achieved through the piezoelectric effect or the photothermal effect[14].

According to the Heisenberg equation of motion and the commutation relations
$[a_{1},a_{1}^{+}]=1$, $[a_{2},a_{2}^{+}]=1$, From Eq.(1), we can get,%
\begin{equation}
\frac{da_{1}}{dt}=-i(\omega _{1}-i\frac{\Gamma _{1}}{2})a_{1}-iJa_{2},
\end{equation}%
\begin{equation}
\frac{da_{2}}{dt}=-i(\omega _{2}-i\frac{\Gamma _{2}}{2})a_{2}-iJa_{1},
\end{equation}%
Here we phenomenologically introduce the mechanical damping rate $%
\Gamma _{j}(j=1,2)$. The resulting equation of motion is%
\begin{equation}
\frac{d}{dt}\binom{a_{1}}{a_{2}}=-i\left(
\begin{array}{cc}
\omega _{1}-i\frac{\Gamma _{1}}{2} & J \\
J & \omega _{2}-i\frac{\Gamma _{2}}{2}%
\end{array}%
\right) \binom{a_{1}}{a_{2}},
\end{equation}%
Then we can introduce the gain and lose through the optomechanical method.
The schematic of our setup is sketched in Fig.1, where two resonators are
optomechanically coupled to two cavities respectively, and simultaneously
coupled to each other. Now we can engineer mechanical gain (loss) by driving
the cavity with a blue-detuned (red-detuned) laser. According to cavity
optomechanics, the optomechanical damping rate is given by [15]%
\begin{equation}
\begin{split}
\gamma _{j}& =g_{0}^{2}n_{cav}^{(j)}\Phi \text{ and } \\
\Phi & =[\frac{-\kappa _{j}}{(\kappa _{j}/2)^{2}+(\Delta _{j}-\omega
_{j})^{2}}+\frac{\kappa _{j}}{(\kappa _{j}/2)^{2}+(\Delta _{j}+\omega
_{j})^{2}}].
\end{split}%
\end{equation}%
where $n_{cav}^{(j)}$ is the intracavity photon number which can be
controlled by the optical drive signal, $\kappa _{j}$ denote the cavity
decay rate, $\Delta _{j}$ represent the laser detuning from the cavity
resonance. The optical frequency shift per displacement is given as $%
G=-\partial \omega _{cav}/\partial x$. For a simple cavity of length $L$, we
have $G=\omega _{cav}/L$, where $\omega _{cav}=\pi (c/L)$ is the frequency
of single optical mode, $c$ is the speed of light in vacuum. And $%
g_{0}=Gx_{ZPF}$ is the vacuum optomechanical coupling strength, expressed as
a frequency, here $x_{ZPF}=\sqrt{\hbar /2m_{j}\omega _{j}}$ is the
zero-point fluctuation amplitude of the mechanical oscillator, and $m_{j}$
is the effective masses of the resonators. Finally, we express $g_{0}$ as a
function of $L$ via%
\begin{equation}
g_{0}=\frac{\pi c}{L^{2}}x_{ZPF}.
\end{equation}%
This indicates that smaller cavities yield larger coupling strengths. Since $%
\gamma _{j}$ can be both positive and negative, it can either increase or
decrease the mechanical damping rate, causes extra damping or antidamping,
corresponding mechanical loss or gain. Thus the total mechanical damping
rates of the resonators can be expressed as the sum of natural mechanical
damping rate $\gamma _{m}$ and the optomechanical damping rate $\gamma _{j}$%
, hence we have $\Gamma _{j}=\gamma _{m}+\gamma _{j}$. Then we consider the
eigenvalues of the effective Hamiltonian in Eq.(4),%
\begin{equation}
\begin{split}
\lambda _{\pm }& =\frac{\omega _{1}+\omega _{2}}{2}-\frac{i}{4}(\Gamma
_{1}+\Gamma _{2})\pm \alpha , \\
\text{with }\alpha & =\sqrt{J^{2}+\frac{1}{4}[(\omega _{1}-\omega _{2})+i(%
\frac{\Gamma _{2}}{2}-\frac{\Gamma _{1}}{2})]^{2}}.
\end{split}%
\end{equation}%
With the simplification, we choose $\Gamma _{1}=-\Gamma _{2}=\Gamma
=-(\gamma _{2}+\gamma _{m})=\gamma _{1}+\gamma _{m}$, we use two identical
optical cavities and resonators, that means $\omega _{j}(j=1,2)=\omega _{m}$%
, $m_{j}(j=1,2)=m$ and $\kappa _{j}(j=1,2)=\kappa $. We also use the driving
lasers with the identical power to drive both micro cavities simultaneously,
this indicates $n_{cav}^{(j)}(j=1,2)=n_{cav}$. Thus Eq. (7) simplifies to%
\begin{equation}
\lambda _{\pm }=\omega _{m}\pm \sqrt{J^{2}-\Gamma ^{2}}.
\end{equation}%
Instead of the traditional vibrational mode, we now have new mechanical
modes, which can be called as the supermodes. Their mechanical frequencies
are $\omega _{\pm }^{(S)}=$Re$(\lambda _{\pm })$ and spectral linewidths are
$\gamma _{\pm }^{(S)}=$ Im$(\lambda _{\pm })$.

\section{MEASUREMENT MECHANISM}

\begin{figure}[tbp]
\includegraphics[width=7.5cm]{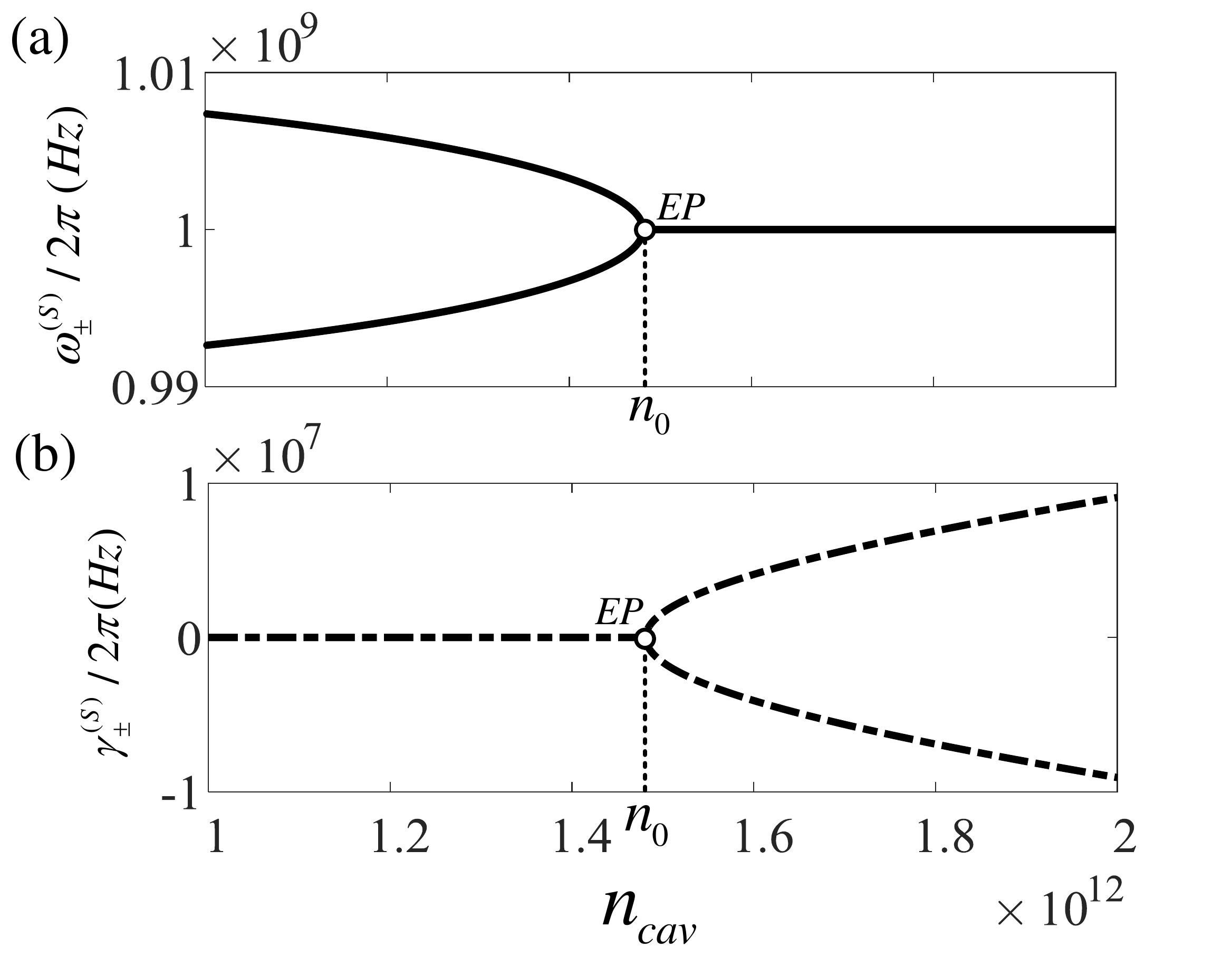}
\caption{ Real(a) and imaginary(b) part of the mechanical eigenfrequencies
as a function of $n_{cav}$. The real part represents the mechanical
supermode frequencies and the imaginary part represents the line widths of the
supermodes.}
\end{figure}
Here we consider the mechanical resonators of Si beams for example, which
possess the thickness of $t=80nm$, mass $m=5.3\times 10^{-3}ng$, the
vibration frequency $\omega _{m}/2\pi =1GHz$[16]. Their mechanical coupling
strength is usually much less than their resonance frequencies[14], here we
use practical device parameter $J/2\pi =10MHz$[17]. We use the cavities with
length $L=0.1mm$ and decay rate $\kappa /2\pi =0.1GHz$. We then set the
cavity-pump detunings $\Delta _{1}=-\Delta _{2}=\omega _{m}$. From Fig.2 we
can see that the parity-time(PT)-broken regime and the PT symmetric regime
are separated by the exception point ($n_{cav}=n_{0}\simeq 1.48\times 10^{12}
$), where the eigenfrequencies coalesce. As $n_{cav}$ increases, the
frequencies of the pair of supermodes approach to each other and coalesce,
while the linewidth starts with zero and then branches, indicating that the
PT symmetry is broken.

Gravitational wave is the propagation of curvature wave in space-time, which
is emitted by accelerated masses. It causes a strain $h$ in space
perpendicular to the direction in which it propagate.
The length change is proportional to the original distance between two
places, $\Delta L/L=h$. We shall see next the splitting of the supermodes
can be used as a signal of the gravitational wave due to the length changes
of cavities.

\begin{figure}[tbp]
\includegraphics[width=8.5cm]{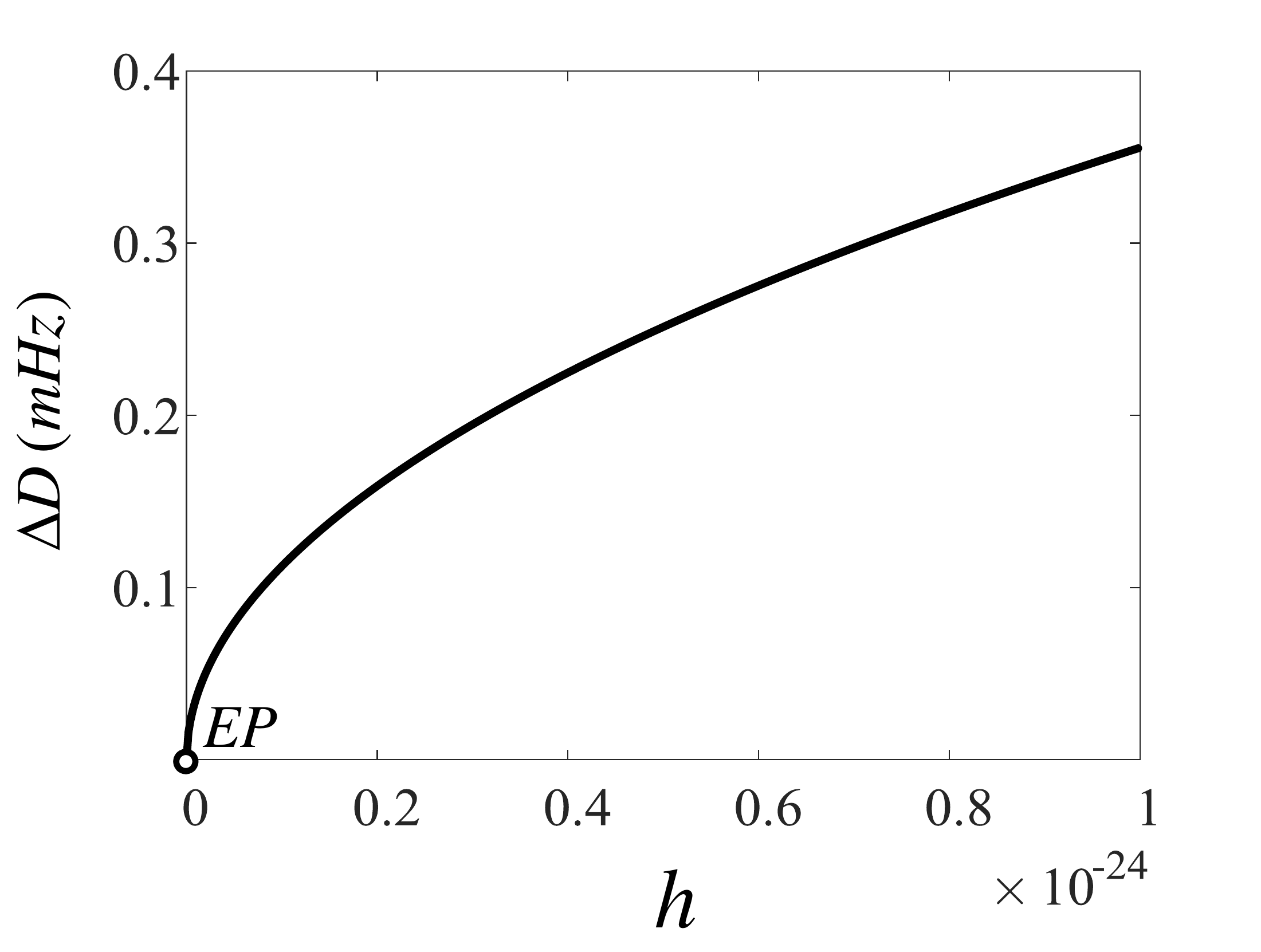}
\caption{ The eigenfrequency splitting as a function of the perturbation strength of the
spatial strain. The frequencies and masses of resonators are $\protect\omega %
_{m}/2\protect\pi =1GHz$ and $m=5.3\times 10^{-3}ng$, respectively. Their
mechanical coupling strength $J/2\protect\pi =10MHz$.}
\end{figure}
An important parameter in a PT symmetrical system is the difference between
the eigenfrequencies, namely the eigenfrequency splitting which can be
defined as $D=\omega _{+}^{(S)}-\omega _{-}^{(S)}$. To evaluate the
sensitivity of gravitational wave detection at EP, we need to know the
effect of a perturbation on the supermode splitting near the EP. Any
perturbation $\Delta L$ of the cavity length induces a change in vacuum
optomechanical coupling strength $dg$ that affects the eigenvalues as
\begin{equation}
D=\text{Re}\left\{ 2\sqrt{J^{2}-[(g_{0}+dg)^{2}n_{0}\Phi ]^{2}}\right\} .
\end{equation}%
Here we consider $\gamma _{m}\ll \gamma _{j}$ and $\Gamma \simeq -\gamma
_{2}=\gamma _{1}$. The relationship between the change of vacuum
optomechanical coupling strength and the length of cavity $\Delta L$ is
given by%
\begin{equation}
dg=\frac{\partial g_{0}}{\partial L}\Delta L=-2\frac{g_{0}}{L}\Delta
L=-2g_{0}h.
\end{equation}%
\ \ The sensitivity can be defined as $\Delta D(h)=D(h)-D(0)$. The system is
at the EP when there is no gravitational wave, thus $D(0)=0$, so we have $%
J=g_{0}{}^{2}n_{0}\left\vert \Phi \right\vert $. Considering $dg\ll g_{0}$,
the result can be simplified as%
\begin{equation}
\Delta D=4\sqrt{2}J\sqrt{h}.
\end{equation}%
Fig.3 shows $\Delta D$ as a function of the perturbation $h$ near the EP. We
can see that the eigenfrequency splitting is proportional to the square root
of the perturbation strength $h$. At the EP, both eigenvalues and
eigenvectors are coalesce. The perturbation of spatial strain can shift the
exceptional point, and thereby the non-Hermitian degeneracy of the
eigenfrequencies are released and cause the supermodes to split. Thus our
scheme does not rely on a shot-noise limited displacement measurement of
mirrors, but rather depends on a precision frequency measurement of the
mechanical mode. The minimum measurable frequency difference is usually
determined by the mechanical linewidth and noises. Here the linewidths of
the mechanical modes are $\gamma _{\pm }^{(S)}=$Im$[\omega _{m}\pm \sqrt{%
J^{2}-[(g_{0}+dg)^{2}n_{0}\Phi ]^{2}}]$. At the EP, $dg=0$, hence $\gamma
_{\pm }^{(S)}=0$, as shown in Fig.2(b). Then we consider the perturbation $%
h>0$, $dg<0$, thereby $J^{2}-[(g_{0}+dg)^{2}n_{0}\Phi ]^{2}>0$, indicates $%
\gamma _{\pm }^{(S)}=0$, the linewidth of the supermodes is still coalesce
and equal to 0, so the perturbation will not induce a linewidth increase.

However, various noise processes will cause the increasing of frequency uncertainty.
For the nanomechanical resonators, the main noise source is the
thermomechanical fluctuations[16]. In such a measurement, the resonator is
driven at a constant mean square amplitude $x_{c}$, which can be roughly
approximated as $x_{c}\approx 0.53t$. According to the
fluctuation-dissipation theorem, the frequency fluctuation induced by
thermal noise can be calculated by $\delta \omega =\sqrt{k_{B}T/2\pi \tau
(m\omega _{m}\left\langle x_{c}^{2}\right\rangle Q)}$. Here $Q$ is the
mechanical quality factor, $T$ is the effective temperature and $k_{B}$ is
the Boltzmann constant. We can see that high mechanical quality and low
temperature help reduce the thermal noise. In order to obtain the
quantum-noise-limited sensitivity of the strain $h$, we assume that the
eigenfrequency split caused by GWs is exactly equal to the frequency
stability determined by the thermal noise, this means $\delta \omega =\Delta
D$, then we get the limit under the sample time of $\tau =1s$.%
\begin{equation}
h_{\min }=\frac{k_{B}T_{eff}}{\left\langle x_{c}^{2}\right\rangle Q}\frac{1}{%
64\pi m\omega _{m}J^{2}}.
\end{equation}%
For the resonator with $Q=10^{5}$, the spatial strain resolution of $h_{\min
}=8.9\times 10^{-25}$ can be achieved at the room temperature ($T_{eff}=300K$%
). If the effective temperature of the vibrational modes can be reduced to $%
1K $ by using the laser cooling technologies[18,19], the sensitivity can be
increased by more than 2 orders of magnitude to reach an unprecedented level
of $h_{\min }=3.0\times 10^{-27}$. We can see from Eq.(12) that the
sensitivity is independent of cavity length used in our scheme, but using a
resonator with higher frequency and smaller mass can improve the detection
sensitivity.

\begin{figure}[tbp]
\includegraphics[width=9cm]{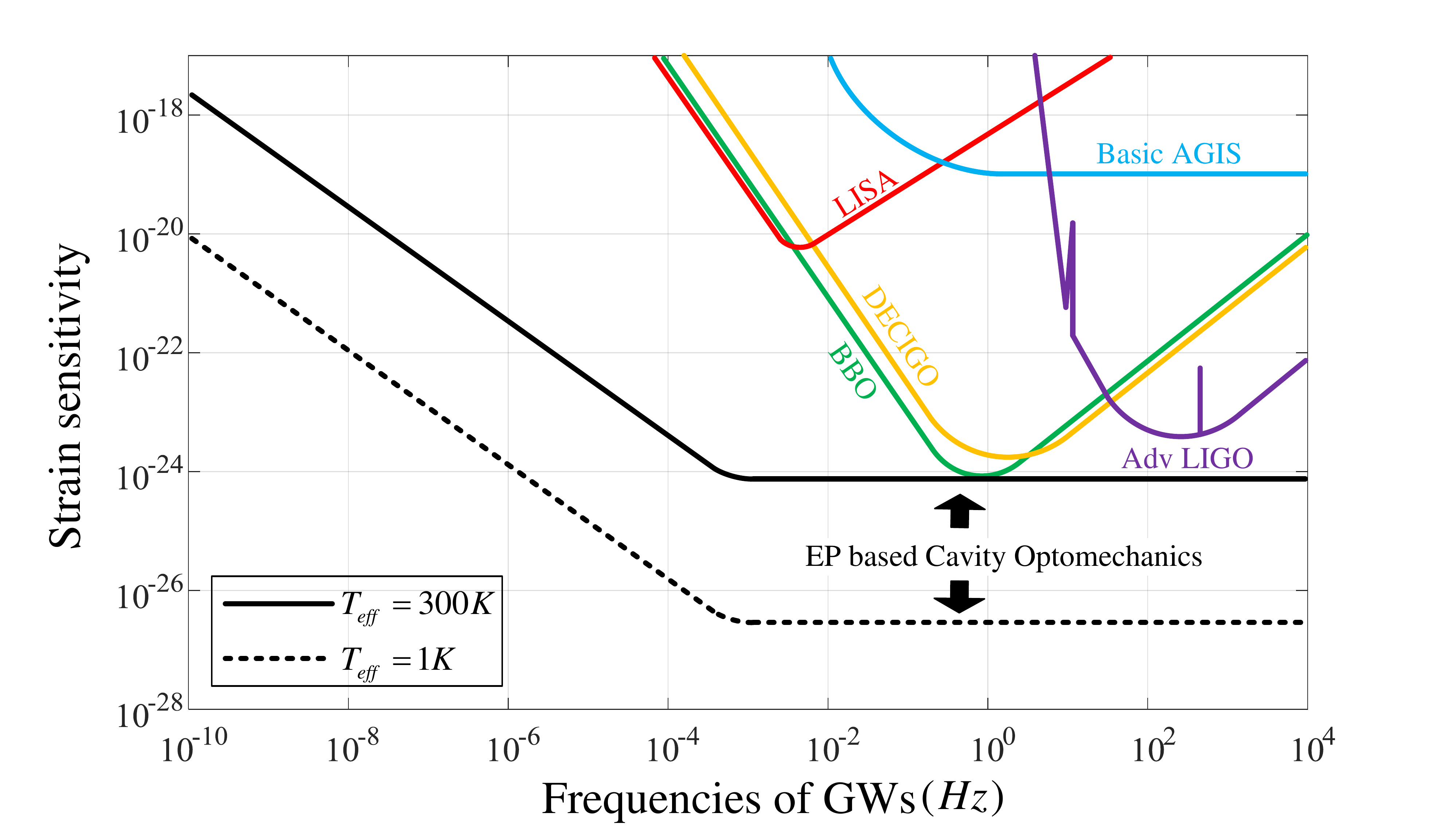}
\caption{ The sensitivity curve of the EP based optomechanical detector with
the maximum observation time of $1h$. The black solid and dashed lines
indicate the theoretical prediction results at effective temperatures of
300K and 1K, respectively. The limit of the advanced LIGO interferometers
and future detectors: LISA, DECIGO, BBO, Basic AGIS, are included for
reference.}
\end{figure}
\

This EP-based optomechanical sensor measures the length changes of the micro
cavities. The spectral splitting is proportional to the square root of the
amplitude of the GW $h$, regardless of the GW's frequency.
Therefore, the detector's responses to gravitational waves of arbitrary
frequency are consistent. However, considering that the distance changes
induced by the ultra-low-frequency gravitational waves are very slow, we
estimate the strain sensitivity with a maximum observation time of $1$ hour,
and compare it with other projects[9-12,20] in Fig.4. It can be clearly seen
that the EP-based optomechanical sensor performs better. We hope that the
proposed system can advance the search for gravitational wave by an enhanced
sensitivity of several orders of magnitude compared with traditional methods
in various frequency bands.

\section{CONCLUSION}

We propose an optomechanical gravitational wave detector based on the
exceptional points. The system is a coupled optomechanical system, in which
the gain and loss are applied by driving the cavities with a blue detuned
and red detuned electromagnetic fields, respectively. When the gain and loss
reach a balance, the system will show the degeneracy of exceptional points,
and the perturbation of the length of micro cavity will cause an
eigenfrequencies split which can be probed with the frequency
spectrum, thus the readout noise caused by the beam amplitude can be
avoided. Compared with the traditional detectors, the sensitivity is greatly
enhanced due to the complex square root topology of EPs in wide frequency
band.

\begin{acknowledgments}
This work was supported by the National Natural Science Foundation of China
(Nos.11274230 and 11574206), the Basic Research Program of the Committee of
Science and Technology of Shanghai (No.14JC1491700).
\end{acknowledgments}

\end{document}